\pdfoutput=1

\documentclass[aps,prl,twocolumn,reprint,superscriptaddress,citeautoscript]{revtex4-2}
\usepackage[dvipdfmx]{graphicx}
\usepackage{subfigmat}
\usepackage[whole]{bxcjkjatype}

\usepackage{amsmath,amssymb}
\usepackage{color}
\usepackage{bm}
\usepackage{physics}
\usepackage{hyperref}
\hypersetup{
	citecolor=blue,
	colorlinks = true,
	urlcolor = blue
}

\usepackage{comment}
\usepackage[normalem]{ulem}

\newcommand{\del}[1]{}

\usepackage{upgreek}

\usepackage{xcolor}
\definecolor{honey}{HTML}{ec9706}

\begin{document}

\title{Fluctuations in Spin Dynamics Excited by Pulsed Light}

\author{Tetsuya Sato}
\affiliation{Institute for Solid State Physics, The University of Tokyo, 5-1-5 Kashiwanoha, Kashiwa, 277-8581 Japan}

\author{Shinichi Watanabe}
\affiliation{Faculty of Science and Technology, Keio University, 3-14-1 Hiyoshi, Kohoku-ku, Yokohama, Kanagawa 223-8522, Japan}

\author{Mamoru Matsuo}
\affiliation{Kavli Institute for Theoretical Sciences, University of Chinese Academy of Sciences, Beijing, China}
\affiliation{CAS Center for Excellence in Topological Quantum Computation, University of Chinese Academy of Sciences, Beijing, China}
\affiliation{Advanced Science Research Center, Japan Atomic Energy Agency, Tokai, Japan}
\affiliation{RIKEN Center for Emergent Matter Science (CEMS), Wako, Saitama, Japan}

\author{Takeo Kato}
\affiliation{Institute for Solid State Physics, The University of Tokyo, 5-1-5 Kashiwanoha, Kashiwa, 277-8581 Japan}

\date{\today}

\begin{abstract}
We theoretically investigate nonequilibrium spin fluctuations in a ferromagnet induced by a light pulse. Using a Lindblad equation consistent with the Landau-Lifshitz-Gilbert equation, we compute the autocorrelation function of magnetization. Our analysis reveals that this function comprises both thermal and nonequilibrium components. To examine the latter in detail, we introduce a Fano factor similar to nonequilibrium current noise in electronic circuits. We demonstrate that this factor encapsulates insights into the transfer of spin units to the environment. Our findings lay the groundwork for nonequilibrium spin noise spectroscopy, offering valuable insights into spin relaxation dynamics.
\end{abstract}
\maketitle 

{\it Introduction.---} The flow of carriers following Poissonian statistics generates current fluctuations, leading to shot noise~\cite{Schottky1918}. Electronic shot noise is proportional to the electron's charge associated with the current. Therefore, the study of shot noise and the determination of the Fano factor have been conducted to understand the nature of charge quanta and interactions between quasi-particles in various correlated electron systems, such as two-dimensional electron gases with fractional charges~\cite{Saminadayar1997, de-Picciotto1997, Reznikov1999, Dolev2008}, metal/superconductor junctions with 2{\it e} charges of Cooper pairs~\cite{Jehl2000, Kozhevnikov2000}, quasi-particle interactions in quantum dots~\cite{Sela2006}, and superconductor-insulator-superconductor tunnel junctions with average charge of Andreev clusters~\cite{Dieleman1997, Jehl1999}. Similarly, the study of spin shot noise is expected to provide insight into the nature of the elementary unit of angular momentum quanta $\hbar$, and to provide information on fundamental spin transport properties~\cite{Chudnovskiy08,Swiebodzinski10,Swiebodzinski2012}. 
Spin shot noise has been theoretically investigated in various magnetic layered systems, including magnetic tunnel junctions~\cite{Foros2005, Swiebodzinski2010}, magnon-mediated spin-transfer systems~\cite{Kamra2016a, Kamra2016b}, low-dimensional systems~\cite{Aftergood18,Joshi18,Aftergood19}, and magnonic thermal-noise systems~\cite{Matsuo2018, Nakata2018}. 
However, spin shot noise has not yet been observed experimentally. All conventional methods for detecting spin shot noise require a specific conversion mechanism from spin noise to electronic noise, and this indirect approach may mask physically meaningful noise in spin systems. Therefore, proposing a direct probe of spin shot noise is crucial.

Optical noise detection provides a direct probe of spin noise in magnetic systems without requiring additional layers for spin detection. Simple transmission and/or Faraday rotation measurements of an optical beam enable spin-noise spectroscopy~\cite{romer2007,Zapasskii:13,Sinitsyn_2016,Smirnov_2021}, which can probe spontaneous spin fluctuations in accordance with the fluctuation-dissipation theorem in paramagnets such as atomic systems~\cite{Aleksandrov1981, Mitsui2000, Crooker2004}, bulk semiconductors~\cite{Oestreich2005, Berski2013}, and semiconductor quantum dots~\cite{Crooker2010, Dahbashi2012}. Spin-noise measurements are also applicable for detecting spontaneous magnon fluctuations in antiferromagnets in a single-crystal form~\cite{Weiss2023}. Nonequilibrium spin-noise spectroscopy has been proposed to study noise properties under an external electric field~\cite{Li2013}, or under optical pumping~\cite{Smirnov2017,Swar2018,Islam2022,Delpy2023}. However, none of the work reported so far suggests the possibility of applying optical noise measurements to detect spin shot noise.

In this Letter, we propose an all-optical approach to detect spin shot noise in a simple thin ferromagnetic film structure. 
The key idea is to use an ultrafast pump laser pulse to impulsively drive the uniform magnetization of a ferromagnet far from equilibrium.
After rapid equilibration of the environment, we can observe a fluctuating magnetization through the Faraday or Kerr rotation signal fluctuations of the probe laser pulse, which reflect nonequilibrium spin relaxation. 
To model this experimental situation, we theoretically calculate the magnon population dynamics using the Lindblad equation and the Fokker-Planck equation, considering the autocorrelation function of the spin component. We analyze the time derivative of the autocorrelation function, which mimics the fluctuation of the spin flow, i.e., the spin current. Here, the flow of spin does not inherently involve spatial diffusion, but represents the energy flow from the spin system to the thermal bath. Therefore, a local optical probe is sufficient to detect the spin current fluctuation. The fluctuation of the spin current can be represented as a sum of two contributions: one is the thermal contribution according to the fluctuation-dissipation theorem, and the other is the contribution of the spin shot noise caused by transferring the quanta of the angular momentum, $\hbar$, between the spin system and the thermal bath. By adopting this approach, the quantitative evaluation of $\hbar$ could be possible by analyzing the noise of the probe pulse in a statistical approach.

\begin{figure}[tb]
    \centering
    \includegraphics[width=85mm]{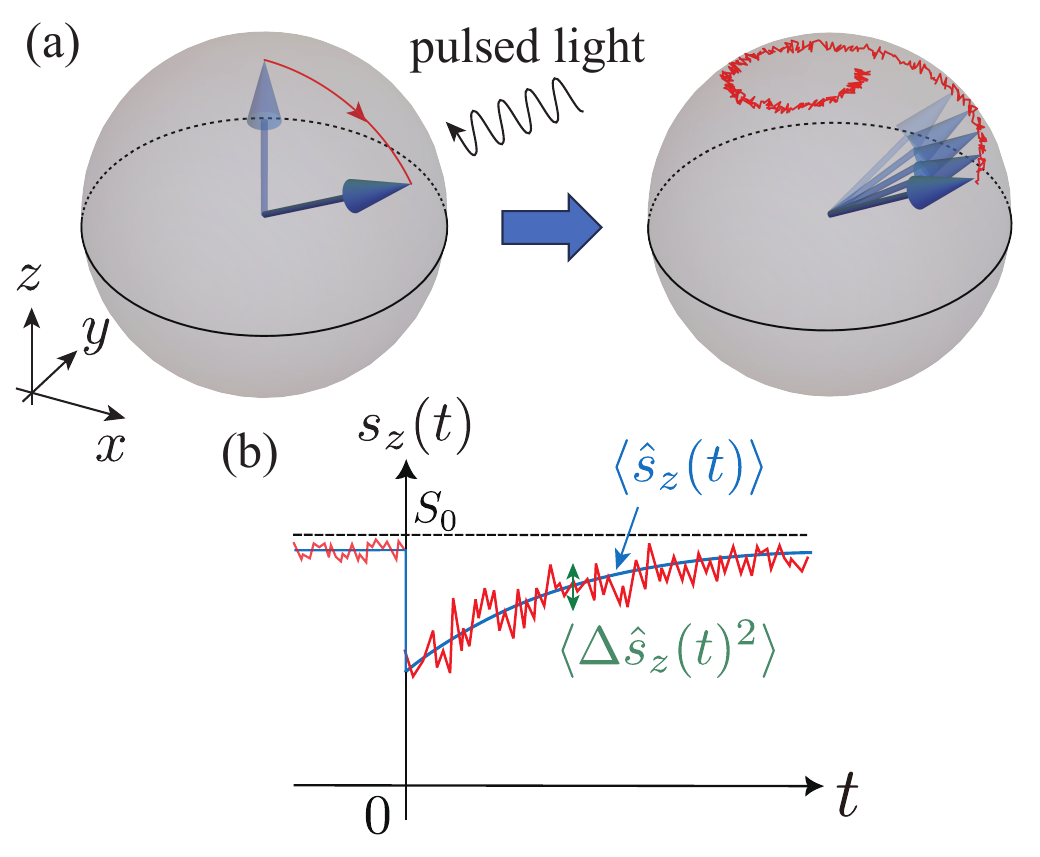}
    \caption{
    (a) A schematic picture for the spin dynamics in our setup. Initially, the spin is in thermal equilibrium and distributed centering on the $z$ axis. After the pulsed light excites the spin of the ferromagnet at $t=0$, the spin precession occurs accompanying the nonequilibrium fluctuation. (b) A schematic picture of the time evolution of $\hat{s}_z$. 
    The red line indicates one measurement of the time evolution of $\hat{s}_z$.
    While the ensemble average with respect to a number of measurements, $\langle \hat{s}_z(t)\rangle$, (the blue line) decays toward the saturated value, $S_0$, there exists nonequilibrium fluctuation represented by $\langle\Delta \hat{s}_z(t)^2\rangle$.}
    \label{fig:setup}
\end{figure}

{\it Setup.---}We examine the dynamics of the total spin $\hat{\bm s}$ or the magnetization $\hat{\bm M} = \hbar \gamma \hat{\bm s}/V$, considering a ferromagnet under a static magnetic field. Here, $\hbar$ represents the Dirac constant, $\gamma\, (<0)$ denotes the gyromagnetic ratio of an electron, and $V$ stands for the volume of the ferromagnetic film. The Hamiltonian is given by
\begin{align}
&{\cal H} = \hbar \gamma \mu_0 {\bm H}_{\rm ex} \cdot \hat{\bm s}, 
\end{align}
where $\mu_0\simeq1.257\times10^{-6}\,{\rm N/A^2}$ denotes the vacuum permeability, and ${\bm H}_{\rm ex} = (0,0,-H_0)$ ($H_0>0$) represents an external magnetic field. Initially, we assume that the averaged spin $\langle {\bm s} \rangle$ in the ferromagnet points in the $+z$ direction and is in thermal equilibrium. We consider spin dynamics induced by pulsed light irradiation at $t=0$. As illustrated in the left panel of Fig.~\ref{fig:setup}(a), the spin state rotates shortly after the laser pulse irradiation due to a sudden change in the direction of the effective magnetic field, induced by laser heating \cite{Kampen2002, Shibata2018}. Subsequently, the direction of the effective magnetic field promptly recovers to the $z$-direction, and after this recovery, spin precession occurs around the $z$-axis, ultimately relaxing into the original thermal equilibrium state due to Gilbert damping (see the right panel of Fig.~\ref{fig:setup}(a)). In our work, we focus on the dynamics and fluctuation of the $z$-component of the spin, as shown in Fig.~\ref{fig:setup}(b).

Throughout this work, we assume that the amplitude of the spin precession is small. Consequently, the Hamiltonian for an isolated spin system can be approximated as $\mathcal{H}=\hbar\omega_0 \hat{b}^\dag \hat{b}+{\rm const.}$, where $\hat{b}$ represents a magnon annihilation operator and $\omega_0=-\gamma \mu_0 H_0$ ($>0$).

{\it Lindblad equation and Fokker-Planck equation.---} To address the quantum nature of the spin system, we introduce the Lindblad equation as follows:
\begin{align}
\label{eq:lindblad}
\dot{\rho}_s(t)=&-\frac{i}{\hbar}[\mathcal{H},\rho_s(t)]+\Gamma \hat{b}\rho_s(t) \hat{b}^\dag-\frac{\Gamma}{2}\{\hat{b}^\dag \hat{b},\rho_s(t)\}\nonumber
\\&+\Gamma' \hat{b}^\dag\rho_s(t) \hat{b} -\frac{\Gamma'}{2}\{\hat{b}\hat{b}^\dag,\rho_s(t)\} \equiv \mathcal{L}\rho_s(t),
\end{align}
where $\rho_s$ denotes the density operator of the spin system. Here, we consider two types of stochastic processes that involve energy exchange between the spin system and the environment: (a) an energy relaxation process, characterized by a transition rate $\Gamma$, and (b) an energy gain process, characterized by a transition rate $\Gamma'$.

The density operator $\rho_s$ can be expressed using the distribution function $P(\beta,\beta^*,t)$ as
\begin{align}
\label{eq:expansion}
\rho_s(t)=\int d^2\beta \ket{\beta}\bra{\beta}P(\beta,\beta^*,t),
\end{align}
where $\ket{\beta}$ represents a coherent state of magnon. Using the distribution function, the Lindblad equation (\ref{eq:lindblad}) is transformed into the following Fokker-Planck equation:
\begin{align}
\label{eq:fokker-planck}
& \partial_t P(\beta,\beta^*,t) = \mathcal{G} P(\beta,\beta^*,t) \nonumber \\ & \equiv 
\left[\left(i\omega_0+\frac{\Gamma-\Gamma'}{2}\right)\partial_\beta\beta+{\rm h.c.} +\Gamma'\partial_\beta\partial_{\beta^*} \right] P(\beta,\beta^*,t).
\end{align}
Solving the Fokker-Plank equation yields the transition probability, defined as the probability of realizing the spin state $\beta'$ at $t+\tau$ starting from a spin state $\beta$ at $t$, as
\begin{align}
\label{eq:transition}
&P(\beta',\beta'^*,t+\tau|\beta,\beta^*,t)=e^{\mathcal{G}\tau}\delta^2(\beta'-\beta)
\nonumber\\&=\mathcal{N}(\tau)^{-1} \exp\left(
-\frac{\pi}{\mathcal{N}(\tau)} \left|\beta'-\beta e^{-\left(\frac{\Gamma-\Gamma'}{2}+i\omega_0\right)\tau}\right|^2
\right),
\end{align}
where $\mathcal{N}(\tau)=\pi\frac{\Gamma'}{\Gamma-\Gamma'}\left(1-e^{-(\Gamma-\Gamma')\tau}\right)$ serves as the normalization factor. It should be noted that, in the limit of $\tau\rightarrow+\infty$, the probability distribution appropriately describes thermal equilibrium, where the spin points to the $z$ axis and fluctuates around it.

{\it Correspondence to the LLG equation.---} The transition rates, $\Gamma$ and $\Gamma'$, can be determined through their correspondence to the Landau-Lifshitz-Gilbert (LLG) equation:
\begin{align}
\label{eq:LLG}
\langle\dot{\hat{{\bm s}}}\rangle =
\langle\hat{{\bm s}}\rangle\times \mu_0\gamma {\bm H}_{\rm ex}-\frac{\alpha}{S_0}\langle\hat{{\bm s}}\rangle\times
\langle\dot{\hat{{\bm s}}}\rangle,
\end{align}
where $\alpha$ represents the Gilbert damping constant. For small amplitudes of spin precession, the LLG equation can be reformulated as
\begin{align}
\left[(1+i\alpha)\partial_t+i\omega_0\right]\langle \hat{b}\rangle=0.
\end{align}
By comparing the solution of the Fokker-Planck equation (\ref{eq:fokker-planck}) with $\alpha \ll 1$, we deduce the condition $\Gamma-\Gamma'=2\alpha\omega_0$. Furthermore, $\Gamma$ and $\Gamma'$ must satisfy the detailed balance condition, $\Gamma = e^{\hbar\omega_0/k_{\rm B}T} \Gamma'$, where $k_B$ denotes the Boltzmann constant. By combining these equations, we derive $\Gamma=2\alpha\omega_0[n_B(\hbar\omega_0)+1]$ and $\Gamma'=2\alpha\omega_0 n_B(\hbar\omega_0)$.

{\it Autocorrelation functions.--} We assume that the spin state is initially prepared as
\begin{align}
\label{eq:initial2}
P(\beta,\beta^*,t=0^+)=\frac{\Gamma-\Gamma'}{\pi\Gamma'}\exp\left(-\frac{\Gamma-\Gamma'}{\Gamma'}\left|\beta-\beta_0\right|^2\right),
\end{align}
where $\beta_0$ denotes the amplitude of the spin excitation from the initial state. The $z$-component of the spin is correlated with the population of the harmonic oscillator as
\begin{align}
& \langle \hat{s}_z(t) \rangle = S_0 - n_{\rm B}(T) - \langle\hat{n}(t)\rangle_{\rm ne} , \\
& \langle\hat{n}(t)\rangle_{\rm ne}=|\beta_0|^2 e^{-(\Gamma-\Gamma')t},
\end{align}
where $\hat{n}=\hat{b}^\dagger \hat{b}$ and $n_{\rm B}(T)=(e^{\hbar\omega_0/k_{\rm B}T}-1)^{-1}$ represents the Bose distribution function. It should be noted that, in the limit of $t\rightarrow \infty$, $\langle\hat{n}(t)\rangle_{\rm ne}$ approaches zero, leaving only the thermal equilibrium part.

The autocorrelation function is defined for $0<t\leq t'$ as
\begin{align}
&C(t',t)\equiv\hbar^2\langle\Delta\hat{s}_z(t')\Delta\hat{s}_z(t)\rangle \\\nonumber&=\hbar^2\Tr\left\{\hat{b}^\dag \hat{b} e^{\mathcal{G}|t'-t|}\left[\hat{b}^\dag \hat{b} \rho_s(t)\right]\right\}  - \hbar^2\langle \hat{b}^\dag \hat{b} (t')\rangle\langle \hat{b}^\dag \hat{b} (t)\rangle ,
\end{align}
whereas $C(t',t)=C(t,t')$ for $0<t'<t$.
Here, $\Delta \hat{s}_z(t) \equiv \hat{s}_z(t)-\langle\hat{s}_z(t)\rangle$ represents the deviation from the average. The autocorrelation can be rewritten using the distribution function as
\begin{align}
C(t',t)&=\hbar^2\int d^2\beta' d^2\beta 
|\beta'|^2
P(\beta',\beta'^*,t'|\beta,\beta^*,t)
\nonumber \\&\times
(\beta^*-\partial_\beta)\beta
P(\beta,\beta^*,t)-\hbar^2\langle \hat{b}^\dag \hat{b} (t')\rangle\langle \hat{b}^\dag \hat{b} (t)\rangle.
\end{align}
After a lengthy but straightforward calculation, the correlation function takes the form
\begin{align}
\label{eq:C}
&C(t',t)= C_{{\rm th}}(t',t)+C_{{\rm ne}}(t',t),
\\
&C_{\rm th}(t',t)= \frac{\hbar^2\Gamma\Gamma'}{(\Gamma-\Gamma')^2}e^{-(\Gamma-\Gamma')|t'-t|}, 
\label{eq:Cth} \\
&C_{{\rm ne}}(t',t) =
\hbar^2\frac{\Gamma+\Gamma'}{\Gamma-\Gamma'}|\beta_0|^2e^{-(\Gamma-\Gamma')t_>} \theta(t) \theta(t'),
\label{eq:Cne} 
\end{align}
where $\theta(t)$ denotes a step function and $t_> = \max(t',t)$. Here, $C_{\rm th}(t',t)$ and $C_{\rm ne}(t',t)$ represent the thermal and nonequilibrium parts of the correlation function, respectively~\footnote{The thermal equilibrium noise can be related to the second-order coherence as $g^{(2)} = \langle \hat{b}^\dag \hat{b}^\dag \hat{b}\hat{b}\rangle/\langle \hat{b}^\dag \hat{b} \rangle^2 = [C_{\rm th}(t,t)/\hbar^2-1]/n_B(T)^2$.
This quantity is calculated as $g^{(2)}=1+\Gamma'/\Gamma$, which takes 2 (1) for high (low) tempetarures.
The result at high temperatures, $g^{(2)}(0)=2$, reflects thermal statistics for bosonic excitations in the ferromagnet, leading to the deviation from Poissonian statistics~\cite{gardiner2004quantum}.
On the other hand, at low temperatures, the noise is suppressed and obeys the Poisson statistics because of the energy gap in the magnon excitation.}.
We observe that only $C_{\rm th}(t',t)$ remains finite in the limit as $|\beta_0|\rightarrow 0$. This implies that our result encompasses the thermal fluctuation of the spin in the ferromagnet in the absence of external light. Particularly, the equal-time correlation function $C_0\equiv C_{\rm th}(t,t)$ is linked to the magnetic susceptibility $\chi = d\langle \hat{M}_z \rangle_{\rm th}/d(-H_0)$ as 
\begin{align}
C_0 = \frac{k_BT V}{\mu_0 \gamma^2} \chi
.
\label{eq:C0}
\end{align}
The fluctuation-dissipation relation suggests that the equal-time correlation function in equilibrium contains the same information as the magnetic susceptibility~\footnote{We can also relate the thermal fluctuation $C_{\rm th}(t',t)$ at different times to a linear response of $\langle \hat{s}_z \rangle$ to a change of the magnetic field. For an explicit discussion of this general fluctuation-dissipation relation, see the Supplemental Material~\cite{Supplement}.}. It is worth noting that this relation can be employed to determine the effective temperature of the environment coupled with the spin system.

{\it Physical interpretation.--} To clarify the physical origin of the nonequilibrium spin fluctuation, we introduce the autocorrelation function \del{of $\dot{\hat{s}}_z$} defined as
\begin{align}
D(t',t)&\equiv \partial_t \partial_{t'} C(t',t).  \del{\hbar^2\langle \Delta\dot{\hat{s}}_z(t')\Delta\dot{\hat{s}}_z(t) \rangle,}
\end{align}
\del{where $\Delta \dot{\hat{s}}_z=\dot{\hat{s}}_z-\langle\dot{\hat{s}}_z\rangle$.} 
This correlation function can be regarded as the fluctuation of the spin flow from the spin system to the thermal bath.
We emphasize that $D(t',t)$ can be obtained experimentally by fitting the measured autocorrelation function $C(t',t)$ and its numerical differentiation\del{, that is, $D(t',t)=\partial_t \partial_{t'} C(t',t)$}.
The correlation function $D(t',t)$ is composed of the delta-function part and the other time-dependent part~\footnote{The nonlocal component $D^c(t',t)$, which is composed of only the thermal part, of the correlation function signifies the breakdown of the assumption of independent angular momentum transfer, that is, the deviation from the Poisson process. Precisely, it emerges from the quantum dynamics of the postselected quantum state following a quantum measurement (quantum jump) accompanied by spin relaxation. Further elaboration on the physical properties of this nonlocal component will be provided elsewhere (also refer to the Supplemental Material~\cite{Supplement}).} as
\begin{align}
D(t',t) = D^0(t) \delta(t'-t) + D^c(t',t).
\label{eq:Dsum}
\end{align}
Note that the delta-function part corresponds to the cusp of $C(t',t)$ at $t=t'$. The equal-time part $D^0(t)$ is calculated as
\begin{align}
D^0(t) &= D^0_{\rm th}+D^0_{\rm ne}(t), \label{eq:D0-1} \\
D^0_{\rm th} &= \frac{2\hbar^2\Gamma\Gamma'}{\Gamma-\Gamma'}, \label{eq:D0-2}\\
D^0_{\rm ne}(t) &=\hbar^2(\Gamma+\Gamma')|\beta_0|^2 e^{-(\Gamma-\Gamma')t}.\label{eq:D0-3}
\end{align}
It is worth noting that $D_{\rm th}^0$ and $D_{\rm ne}^0(t)$ originate from $C_{\rm th}(t',t)$ and $C_{{\rm ne}}(t',t)$, respectively.

\begin{figure}[tb]
    \centering
    \includegraphics[width=90mm]{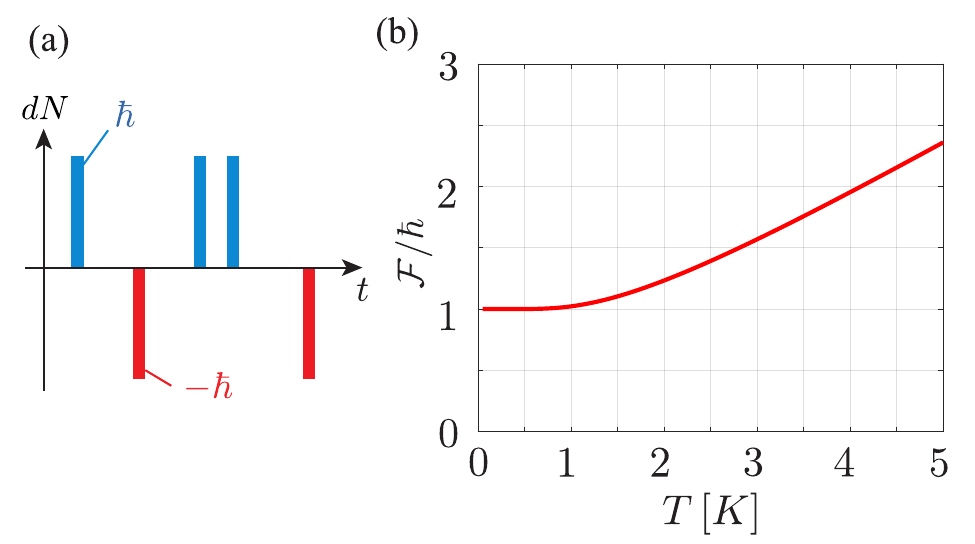}
    \caption{(a) A schematic illustration of the stochastic dynamics of the change of the spin, $dN$ \del{$\dot{s}_z$}. 
    (b) The Fano factor, which is analytically calculated as $\mathcal{F}=\coth\left(\hbar\omega_0/2k_BT\right)\hbar$ (see Eq.~(\ref{eq:Fanodef})), is plotted as a function of temperature. At low temperatures, $\mathcal{F}/\hbar$ approaches unity.
    }
    \label{fig:poisson}
\end{figure}


A simple physical interpretation is possible when the nonequilibrium part is dominant ($D(t',t)\simeq D^0_{\rm ne}(t)\delta(t'-t)$).
This condition is met for a certain duration after applying a sufficiently strong pulse to the system, i.e., when $n_{\rm B}(T) \ll |\beta_0|^2 e^{-(\Gamma-\Gamma')t}$~\footnote{We consider that the amplitude of the excitation induced by the laser pulse irradiation is much larger than that of the thermal fluctuation. This is a usual condition for which the ultra-fast time-resolved spin precession is observed.}. 
We then define the Fano factor as the ratio between the spin flow and its fluctuation, denoted by $\mathcal{F} = D^0_{\rm ne}(t)/ \partial_t \langle \hat{s}_z(t)\rangle \del{\langle \dot{\hat{s}}_z(t) \rangle}$. Using Eq.~(\ref{eq:D0-3}), this factor is calculated as
\begin{align}
\mathcal{F} =
\frac{\Gamma+\Gamma'}{\Gamma-\Gamma'}\hbar = 
\coth\left(\frac{\hbar\omega_0}{2k_BT}\right)\hbar .
\label{eq:Fanodef}
\end{align}
The result (\ref{eq:Fanodef}) provides crucial insights into the angular momentum transfer associated with a single relaxation process in a ferromagnet, particularly under conditions of low temperature~\footnote{Eq.~(\ref{eq:Fanodef}) is obtained assuming that the change of the spin due to the coupling to the environment is $\pm\hbar$. If there exists two-magnon relaxation (excitation) process due to the environment, for example, the Fano factor is modified by mixing of the two spin relaxation (excitation) processes with the spin change being $\pm\hbar$ and $\pm 2\hbar$.}. To illustrate this, we demonstrate how identical conclusions can be drawn from a straightforward analysis based on the Poisson process. 
In Fig.~\ref{fig:poisson}(a), we schematically depict the stochastic dynamics of the change of the spin, $dN$ \del{$\dot{\hat{s}}_z$} in the range of $[t,t+\tau]$, where $\tau$ is chosen small enough not to alter the amplitude of the spin precession, $|\langle \hat{b}(t) \rangle|$. Each spin transfer event changes the magnitude of the spin \del{$\hat{s}_z$} by $\pm \hbar$, thus the dynamics of $dN/dt$ \del{$\dot{\hat{s}}_z$} can be described by a series of delta functions, with weights of $\hbar$ or $-\hbar$. If these transfer events are stochastically independent, the probability of observing $n$ events of $\hbar$ spin transfer and $m$ events of $-\hbar$ spin transfer during the time interval $[t,t+\tau]$ is expressed as $P_n P_m$. Here, $P_n$ and $P_m$ follow the Poisson distribution with averages ${\rm E}[n] \del{\langle n \rangle} = \lambda \Gamma\tau$ and ${\rm E}[m] \del{\langle m \rangle} = \lambda \Gamma'\tau$, respectively, where $\lambda$ depends on the amplitude of the spin precession and ${\rm E}[\cdot]$ denotes a classical expectation value. 
Using properties of the Poisson distribution, we calculate the average and variance of $dN/dt$ \del{$\dot{\hat{s}}_z$} as ${\rm E}[dN/dt]=\hbar({\rm E}[n]-{\rm E}[m])/\tau \del{\langle \dot{\hat{s}}_z\rangle =\hbar (\langle n \rangle-\langle m \rangle)/\tau}=\lambda\hbar(\Gamma-\Gamma')$ and ${\rm E}[(dN/dt)^2]=\hbar^2({\rm E}[n]+{\rm E}[m])/\tau \del{\langle (\Delta \dot{\hat{s}}_z)^2\rangle = \hbar^2 [\langle (\Delta n)^2 \rangle+\langle (\Delta m)^2 \rangle]/\tau} = \lambda \hbar^2(\Gamma+\Gamma')$. 
Considering that ${\rm E}[dN/dt]$ and ${\rm E}[(dN/dt)^2]$ correspond to $\partial_t\langle \hat{s}_z\rangle$ and $D_{\rm ne}^0$, respectively, this calculation leads to the result (\ref{eq:Fanodef}), affirming the efficacy of the current intuitive discussion.
We note that this interpretation based on the Poissonian process holds only under the nonequilibrium condition ($D(t',t)\simeq D^0_{\rm ne}(t)\delta(t'-t)$). 
When the system is near the thermal equilibrium, each relaxation process is not independent due to the existence of the nonlocal part $D^c(t',t)$ (see Supplemental Material~\cite{Supplement}).

We show the temperature dependence of the Fano factor in Fig.~\ref{fig:poisson}(b) for $\mu_0 H_0 = 3\, {\rm T}$ and $\gamma=-1.97\times 10^{11}\,{\rm rad/(s\cdot T)}$ ($\omega_0/2\pi = 94\,{\rm GHz}$)~\cite{Yin2015}.
This dependence can be elucidated as follows.
At low temperatures ($k_BT\ll \hbar\omega_0$), only the energy relaxation process predominates ($\Gamma \gg \Gamma'$), resulting in the Fano factor becoming $\hbar$. Through an intuitive discussion grounded in the Poisson process, this outcome signifies the transfer of angular momentum from the spin system to the bath in units of $\hbar$. It is worth noting that a similar rationale has been applied in determining the unit of charge in electronic transport, derived from nonequilibrium current noise (shot noise).
As the temperature increases, the energy gain process also becomes significant. Consequently, the presence of two distinct transition processes diminishes the average spin flow, although it contributes additively to its fluctuation. Consequently, the Fano factor begins to increase with increasing temperature.

\begin{figure}[tb]
    \centering
    \includegraphics[width=80mm]{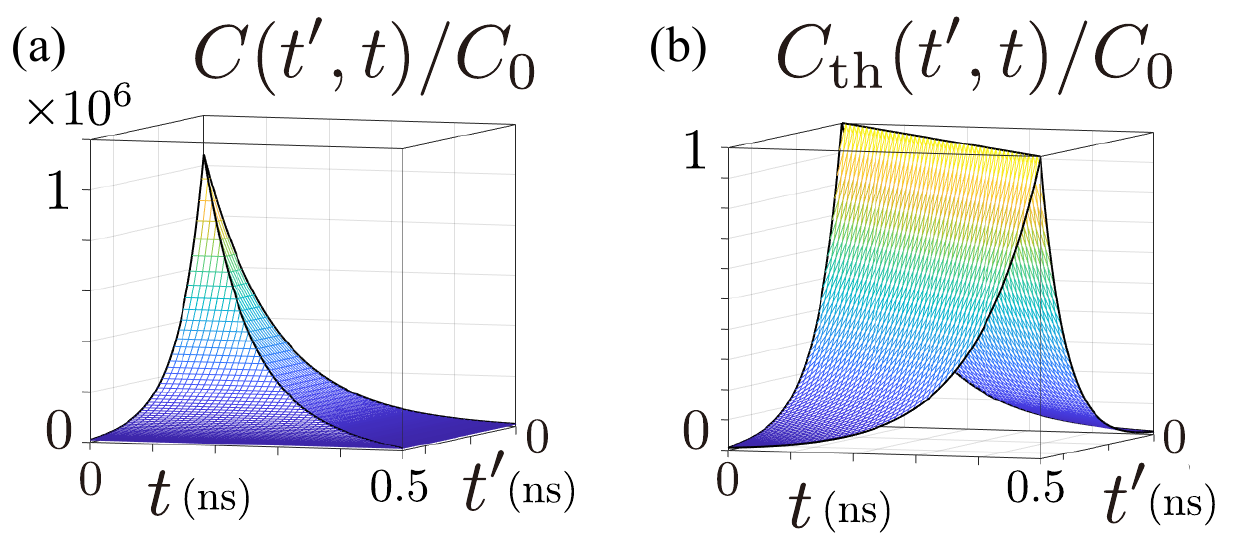}
    \caption{ 
    The correlation function of $\Delta\hat{s}_z$ is shown as functions of $t'$ and $t$. (a) the total fluctuation $C(t',t)/C_0$, (b) the thermal part $C_{\rm th}(t',t)/C_0$. 
    Here $C_0\approx 4.4\times10^{3} \hbar^2$ is the equal-time thermal fluctuation.
    }
    \label{fig:SzCorrelation}
\end{figure}

{\it Experimental protocol.--} Finally, we outline a feasible experimental protocol to observe the Fano factor. We propose using a thin film of ferromagnetic permalloy as the sample.
Specific experimental conditions underlying our calculation are described in Supplemental Material~\cite{Supplement}.
Ultrafast dynamics of the $z$-component of the sample's magnetization, $M_z(t)$, is measured following laser pulse irradiation, employing time-resolved magneto-optical Kerr effect measurements~\cite{Kampen2002, Shibata2018}. 
We can deduce the ensemble average of the spin flow, $\partial_t\langle\hat{s}_z(t)\rangle$ \del{$\langle \dot{\hat{s}}_z(t) \rangle$}, from the time derivative of the average across trials of the magnetization dynamics measurements, i.e., $d\langle M_z(t) \rangle/dt$. Then, the correlation function $C(t',t)$ is obtained by calculating the covariance of $M_z$ at different times, that is, $\langle M_z(t) M_z(t') \rangle$.
As an illustrative example, we show the autocorrelation function for a $1\, \mu{\rm m}\times 1 \, \mu{\rm m}\times 5 \, {\rm nm}$ ferromagnetic film in Fig.~\ref{fig:SzCorrelation}(a).
The parameters are as follows: $\mu_0 H_0= 3\,{\rm T}$, $\gamma=-1.97\times 10^{11}\,{\rm rad/(s\cdot T)}$, $T=300\,{\rm K}$, $\beta_0=(S_0/5)^{1/2}$, $\alpha=7.8\times10^{-3}$, together with saturation magnetization of $\mu_0 M_0 = 0.905\,{\rm T}$~\cite{Yin2015}.
We also show the thermal part $C_{\rm th}(t',t)$ separately in Fig.~\ref{fig:SzCorrelation}(b).
As indicated by the comparison between Fig.~\ref{fig:SzCorrelation}(a) and (b), the nonequilibrium part $C_{\rm ne}(t',t)$ always dominates the thermal part $C_{\rm th}(t',t)$ in the time range shown here.
From a cusp in $C(t',t)$ along the line $t'=t$, we can obtain the delta function part of the correlation function, $D^0_{\rm ne}(t)$ (see Eq.~(\ref{eq:Dsum})), which can be used to obtain the Fano factor, combining the time derivative of the mean magnetization value. Furthermore, as illustrated in Fig. ~\ref{fig:poisson} (b), we can quantitatively determine the elementary unit of angular momentum, $\hbar$, from the Fano factor. The conditions required for this experiment---low temperature ($<$ 1 K) and high magnetic field ($\sim$ 3 T)---are experimentally feasible.

{\it Summary.--} We investigated the nonequilibrium fluctuation arising from the ferromagnetic magnetization under pulse irradiation. 
We calculated the Fano factor, which is defined as the ratio between the nonequilibrium spin current flowing out of the spin system and its nonequilibrium fluctuation and observed that the Fano factor measured at low temperature offers insight into the unit of angular momentum transferred per spin relaxation process in a bulk ferromagnet. 
Our proposal sets the stage for nonequilibrium spin-noise spectroscopy, offering an advanced technique to access information that is inaccessible by other experimental means.

\begin{acknowledgments}
We thank JSPS KAKENHI for Grants (No.~23KJ0702, No.~23H01839, No.~24H00322, and No.~24K06951). 
M.M.~is partially supported by the National Natural Science Foundation of China (NSFC) under Grant No.~12374126 and the Priority Program of the Chinese Academy of Sciences, Grant No.~XDB28000000.
\end{acknowledgments}

\bibliography{ref}

\end{document}


\title{Supplementary Information: Fluctuations in spin dynamics excited by pulsed light}

\author{Tetsuya Sato}
\affiliation{Institute for Solid State Physics, The University of Tokyo, 5-1-5 Kashiwanoha, Kashiwa, 277-8581 Japan}

\author{Shinichi Watanabe}
\affiliation{Faculty of Science and Technology, Keio University, 3-14-1 Hiyoshi, Kohoku-ku, Yokohama, Kanagawa 223-8522, Japan}

\author{Mamoru Matsuo}
\affiliation{Kavli Institute for Theoretical Sciences, University of Chinese Academy of Sciences, Beijing, China}
\affiliation{CAS Center for Excellence in Topological Quantum Computation, University of Chinese Academy of Sciences, Beijing, China}
\affiliation{Advanced Science Research Center, Japan Atomic Energy Agency, Tokai, Japan}
\affiliation{RIKEN Center for Emergent Matter Science (CEMS), Wako, Saitama, Japan}

\author{Takeo Kato}
\affiliation{Institute for Solid State Physics, The University of Tokyo, 5-1-5 Kashiwanoha, Kashiwa, 277-8581 Japan}

\date{\today}

\maketitle

\section{Derivation of $P(\beta,\beta^*,t=0^+)$}
\label{app:pulse}

In the main text, the initial distribution function, $P(\beta,\beta^*,t=0^+)$, is assumed to be of the form of Eq.~(8).
There are a few ways to prepare this initial spin state.
As a simple example, let us consider the case that the external pulse light changes the spin state of the ferromagnet through the Zeeman field, whose Hamiltonian is given as
\begin{align}
\mathcal{H}_{\rm pulse}=-i\hbar(\hat{b}-\hat{b}^\dag)F_0\delta(t),
\end{align}
where a real constant $F_0$ indicates the magnitude of the pulse.
The density operator in pulse irradiation evolves as
\begin{align}
\frac{d\rho_s(t)}{dt} = -\frac{i}{\hbar}[\mathcal{H}_{\rm pulse}^{0},\rho_s(t)].
\end{align}
Using the expression 
\begin{align}
\label{eq:expansion}
\rho_s(t)=\int d^2\beta \ket{\beta}\bra{\beta}P(\beta,\beta^*,t),
\end{align}
we obtain
\begin{align}
\partial_t P(\beta,\beta^*,t) = - F_0\delta(t)(\partial_\beta + \partial_{\beta^*})P(\beta,\beta^*,t).
\end{align}
We assume that the distribution function just before the pulse irradiation is given by the thermal equilibrium distribution, that is,
\begin{align}
P_{\rm eq}=\mathcal{N}^{-1} \exp\left(-\frac{\Gamma-\Gamma'}{\Gamma'}|\beta|^2\right), 
\end{align}
where $\mathcal{N}=\pi\Gamma'/(\Gamma-\Gamma')$.
Then, the distribution function just after pulse irradiation, $P(\beta,\beta^*,t=0^+)$, is obtained as
\begin{align}
P(\beta,\beta^*,t=0^+)  \nonumber
&= \exp\left(-\int_{0^-}^{0^+} dt F_0\delta(t)(\partial_\beta+\partial_{\beta^*}) \right)P_{\rm eq}
\\\nonumber&= \mathcal{N}^{-1} \sum_{n,m,l} \frac{(-F_0)^n}{n!}\frac{(-F_0)^m}{m!}\frac{\left(-\frac{\Gamma-\Gamma'}{\Gamma'}\right)^l}{l!}\partial_\beta^n\partial_{\beta^*}^m \beta^{l}\beta^{*^l}
\\\nonumber&=\mathcal{N}^{-1}\sum_{n,m,l}{}_lC_n {}_lC_m (-F_0)^n\beta^{l-n}(-F_0)^m\beta^{*^{l-m}}\frac{\left(-\frac{\Gamma-\Gamma'}{\Gamma'}\right)^l}{l!}
\\
&=\mathcal{N}^{-1}\exp\left(-\frac{\Gamma-\Gamma'}{\Gamma'}|\beta-F_0|^2\right).
\end{align}
Thus, we can derive Eq.~(8) in the main text.

\section{Fluctuation-Dissipation Theorem}
\label{app:FDtheorem}

In this appendix, we derive the fluctuation-dissipation theorem which connects $C_{\rm th}(t',t)$, the thermal auto-correlation function of $\hat{s}_z$, and the linear response of $\langle \hat{s}_z \rangle$ against a perturbation of the magnetic field.
From now on, we denote a time variable with $\tau$ for convenience of our discussion.
We consider the situation that the system is in the thermal equilibrium before $\tau=0$ with the Hamiltonian, $\mathcal{H}_{\tau<0}=\hbar\omega_0\hat{b}^\dag\hat{b}$, and after $\tau=0$ the magnetic field is perturbed as 
\begin{align}
    \mathcal{H}_{\tau>0} = \hbar(\omega_0+\epsilon)\hat{b}^\dag\hat{b}.
\end{align}
Here, we assume that the perturbation is temporally uniform and that $\epsilon$ is constant.
The damping parameters, $\Gamma$ and $\Gamma'$, also deviate after the perturbation at $\tau=0$ as
\begin{align}
&\Gamma_{\tau>0} = 2\alpha(\omega_0+\epsilon)[n_B(\hbar(\omega_0+\epsilon))+1]\simeq \Gamma + 2\alpha\epsilon \left[ \frac{e^{\beta\hbar\omega_0}}{e^{\beta\hbar\omega_0}-1} -  \frac{\beta\hbar \omega_0 e^{\beta\hbar\omega_0}}{(e^{\beta\hbar\omega_0}-1)^2}\right]
\\
&\Gamma'_{\tau>0} = 2\alpha(\omega_0+\epsilon) n_B(\hbar(\omega_0+\epsilon)) \simeq \Gamma' + 2\alpha \epsilon \left[\frac{1}{e^{\beta\hbar\omega_0}-1} - \frac{\beta\hbar\omega_0 e^{\beta\hbar\omega_0}}{(e^{\beta\hbar\omega_0}-1)^2} \right] 
\end{align}
Then, the Fokker-Planck equation of $\tau>0$ is given as
\begin{align}
\label{app:eq:Fokker-Planck}
\partial_\tau P_{\tau>0}(\beta,\beta^*,\tau) = \left[\left(i\omega_0+i\epsilon+\frac{\Gamma_{\tau>0}-\Gamma'_{\tau>0}}{2}\right)\partial_\beta\beta+{\rm h.c.} +\Gamma'_{\tau>0}\partial_\beta\partial_{\beta^*} \right] P_{\tau>0}(\beta,\beta^*,\tau).
\end{align}
Since the distribution function of $\tau<0$ coincides with that in equilibrium, $P_{\rm eq}(\beta,\beta^*)\equiv \frac{\Gamma-\Gamma'}{\pi\Gamma'}\exp\left(-\frac{\Gamma-\Gamma'}{\Gamma'}\left|\beta\right|^2\right)$, we can expand the distribution function as $P_{\tau>0}(\beta,\beta^*\tau)=P_{\rm eq}(\beta,\beta^*) + \epsilon P_1(\beta,\beta^*,\tau)$ up to the first order of $\epsilon$.
We note $\int d^2\beta P_1(\beta,\beta^*,\tau)=0$ since the distribution function is normalized.
Following the Fokker-Planck equation (\ref{app:eq:Fokker-Planck}), the deviated distribution function, $P_1(\beta,\beta^*,\tau)$, temporally evolves as
\begin{align}
\partial_\tau P_1(\beta,\beta^*,\tau) = &\left[\left(i\omega_0+\frac{\Gamma-\Gamma'}{2}\right)\partial_\beta\beta+{\rm h.c.} +\Gamma'\partial_\beta\partial_{\beta^*} \right] P_1(\beta,\beta^*,\tau) 
\\\nonumber& + \left[\left(i+\alpha \right) \partial_\beta \beta +h.c.  +2\alpha  \left( \frac{1}{e^{\beta\hbar\omega_0}-1} - \frac{\beta\hbar\omega_0 e^{\beta\hbar\omega_0}}{(e^{\beta\hbar\omega_0}-1)^2}\right)\partial_\beta\partial_{\beta^*} \right]P_{\rm eq}(\beta,\beta^*) .
\end{align}
By multiplying $\epsilon|\beta|^2$ and integrating about $\beta$, we obtain the time evolution of $\Delta\langle\hat{b}^\dag\hat{b}\rangle(\tau)\equiv \langle\hat{b}^\dag\hat{b}\rangle(\tau)-\langle\hat{b}^\dag\hat{b}\rangle_{\rm th}$ up to the first order of $\epsilon$ as
\begin{align}
\partial_\tau \Delta\langle \hat{b}^\dag\hat{b} \rangle(\tau) = -2\alpha\omega_0\Delta\langle\hat{b}^\dag\hat{b}\rangle (\tau)- 2\alpha\epsilon \langle\hat{b}^\dag\hat{b}\rangle_{\rm th} + 2\alpha\epsilon \left( \frac{1}{e^{\beta\hbar\omega_0}-1} - \frac{\beta\hbar\omega_0 e^{\beta\hbar\omega_0}}{(e^{\beta\hbar\omega_0}-1)^2}\right),
\end{align}
where $\tau>0$ and we used $\Gamma-\Gamma'=2\alpha\omega_0$.
We can analytically solve the above equation, and obtain the linear response of $\hat{b}^\dag\hat{b}$ given as
\begin{align}
\label{app:eq:sz}
\Delta\langle\hat{b}^\dag\hat{b}\rangle(\tau) &= \left[ - \frac{\epsilon}{\omega_0} \langle\hat{b}^\dag\hat{b}\rangle_{\rm th} + \frac{\epsilon}{\omega_0} \left( \frac{1}{e^{\beta\hbar\omega_0}-1} - \frac{\beta\hbar\omega_0 e^{\beta\hbar\omega_0}}{(e^{\beta\hbar\omega_0}-1)^2}\right)\right] \left(1-e^{-2\alpha\omega_0\tau}\right)
\\\nonumber &=   - \frac{\epsilon}{\omega_0} \frac{\beta\hbar\omega_0 e^{\beta\hbar\omega_0}}{(e^{\beta\hbar\omega_0}-1)^2} \left(1-e^{-2\alpha\omega_0\tau}\right),
\end{align}
where we used $\langle\hat{b}^\dag\hat{b}\rangle_{\rm th} = 1/(e^{\beta\hbar\omega_0}-1)$.
Considering that the thermal auto-correlation function of $\hat{s}_z$ is given as (see Eq.~(14) in the main article)
\begin{align}
C_{\rm th}(t+\tau,t)= \frac{\hbar^2\Gamma\Gamma'}{(\Gamma-\Gamma')^2}e^{-(\Gamma-\Gamma')\tau}=\frac{\hbar^2 e^{\beta\hbar\omega_0}}{\left(e^{\beta\hbar\omega_0}-1\right)^2}e^{-2\alpha\omega_0\tau} \,\,\,\,\, (\tau>0),
\end{align}
we can finally derive the fluctuation-dissipation theorem as 
\begin{align}
\label{app:eq:FD}
\langle\hat{b}^\dag\hat{b}\rangle(\tau)-\langle\hat{b}^\dag\hat{b}\rangle_{\rm th} = - \epsilon\frac{\beta}{\hbar} \left[C_{\rm th}(t,t)-C_{\rm th}(t+\tau,t)\right]. 
\end{align}
When enough time passes ($\tau\rightarrow\infty$) and the system reaches a new equilibrium of $\mathcal{H}_{\tau>0}$, the left hand side of Eq.~(\ref{app:eq:FD}) can be rewritten as 
\begin{align}
\langle\hat{b}^\dag\hat{b}\rangle(\tau\rightarrow\infty)-\langle\hat{b}^\dag\hat{b}\rangle_{\rm th} = \epsilon \frac{d\langle\hat{b}^\dag\hat{b}\rangle_{\rm th}}{d\omega_0} = \frac{\epsilon}{\mu_0\gamma} \frac{d\langle\hat{b}^\dag\hat{b}\rangle_{\rm th}}{d(-H_0)} .
\end{align}
By introducing the relation between the magnetization and the magnon population, $\hat{M}_z=\hbar\gamma (S_0-\hat{b}^\dag\hat{b})/V$, and by defining the magnetic susceptibility as $\chi\equiv d\langle\hat{M}_z\rangle_{\tau\rightarrow\infty}/d(-H_0)$, we can obtain Eq.~(16) in the main article:
\begin{align}
C_{\rm th}(t,t)=\frac{k_BTV}{\mu_0\gamma^2}\chi.
\end{align}

\section{Calculation of $D(t',t)$}
\label{app:S}

As seen in Eq.~(18) in the main text, the correlation function of $D(t',t)$ is given by the sum of the delta function part $D^0(t)\delta(t-t')$
and the nonlocal part $D^c(t',t)$.
The nonlocal component $D^c(t',t)$ can be expressed for $t\neq t'$ as
\begin{align}
\label{eq:back}
D^c(t',t) =
\partial_t\partial_{t'}C(t',t)
=-\hbar^2\Gamma\Gamma'e^{-(\Gamma-\Gamma')|t'-t|}.
\end{align}
On the other hand, the equal-time component $D^0(t)\delta(t-t')$ is produced by a cusp at $t'=t$ in $C(t,t')$.
Since the integral of $D(t',t)$ near $t'=t$ is calculated as
\begin{align}
\nonumber
\lim_{\epsilon\rightarrow0}\int_{t-\epsilon}^{t+\epsilon} dt' D(t',t) &= \lim_{\epsilon\rightarrow0} \int_{t-\epsilon}^{t+\epsilon} dt' \partial_{t'}\partial_{t}C(t',t)
\\\nonumber &=\left.\partial_t C(t',t)\right|_{t'=t+0}-\left.\partial_t C(t',t)\right|_{t'=t-0}
\\
&=\frac{2\hbar^2\Gamma\Gamma'}{\Gamma-\Gamma'}+\hbar^2(\Gamma+\Gamma') |\beta_0|^2 e^{-(\Gamma-\Gamma')t},
\label{eq:app_integral}
\end{align}
we obtain
\begin{align} 
D_0(t) = \hbar^2\left[\frac{2\Gamma\Gamma'}{\Gamma-\Gamma'}+(\Gamma+\Gamma')|\beta_0|^2 e^{-(\Gamma-\Gamma')t}\right].
\label{eq:shot}
\end{align}

\section{Stochastic Representation}
\label{app:stochastic}

In the main text, \del{the fluctuation of $\dot{\hat{s}}_z$, i.e., }$D(t',t)$ is calculated by differentiating the correlation function for $\hat{s}_z$, i.e., $C(t',t)$ with respect to time.
As mentioned in the main text, $D(t',t)$ is understood from the stochastic representation.
In this appendix, we derive the same result for $D(t',t)$ using the formulation for the stochastic process~\cite{breuer2002theory,gardiner2004quantum,wiseman2009quantum} to obtain a physical picture for each term in $D(t',t)$.

We consider the Poissonian process between time $t$ and time $t+dt$. \del{In this minute time range, $\dot{\hat{s}}_z$ takes only three values, $0$ and $\pm1$.} We define the random variables $dN_+$ ($dN_-$), which takes $1$ when the energy relaxation (the energy gain) occurs, and which takes zero in other cases. 
By defining $\hat{L}_+=\sqrt{\Gamma}\hat{b}$ and $\hat
{L}_-=\sqrt{\Gamma'}\hat{b}^\dag$, these random variables satisfies the following relations:
\begin{align}
\label{eq:app_poisson}
    &dN_l(t)^2=dN_l(t),\\
&dN_+(t)dN_-(t)=0,\label{eq:app_poisson2}\\
&{\rm E}[dN_l(t)] \del{\langle dN_l(t)\rangle} = dt \langle \hat{L}_l^\dag \hat{L}_l(t)\rangle,
\end{align}
where $l$ takes $\pm$ and Eq.~(\ref{eq:app_poisson2}) indicates that $dN_+$ process and $dN_-$ process do not occur at the same time.

Using this stochastic representation, we can define a spin loss rate, $I_s(t)$, as \del{redefine $\dot{\hat{s}}_z$ as}
\begin{align}
I_s(t) \del{\dot{\hat{s}}_z}=\frac{dN_+}{dt}-\frac{dN_-}{dt}=\sum_{l=\pm} l \frac{dN_l}{dt}.
\end{align}
Using this notation, the correlation function of $\Delta I_s\equiv I_s-{\rm E}[I_s]$ is calculated as \del{$D(t',t)$ can be rewritten as}
\begin{align}
\label{eq:app_s}
\del{D(t',t)} \hbar^2{\rm E}[\Delta I_s(t')\Delta I_s(t)] &=\frac{\hbar^2}{dt^2}\sum_{l,l'=\pm}l'l\Bigl\{{\rm E} [dN_{l'}(t') dN_{l}(t)] - {\rm E}[dN_{l'}(t')]{\rm E}[dN_{l}(t)]\Bigr\}.  \del{\Bigl\{\langle dN_{l'}(t') dN_{l}(t)\rangle -\langle dN_{l'}(t')\rangle\langle dN_{l}(t)\rangle\Bigr\}.}
\end{align}
The delta-function part \del{in $D(t',t)$} is calculated using Eq.~(\ref{eq:app_poisson}) and focusing on the infinitesimal region near $t'=t$ as
\begin{align}
\del{D(t',t)} \hbar^2{\rm E}[\Delta I_s(t')\Delta I_s(t)] &=\frac{\hbar^2}{dt^2}\left\{{\rm E} [dN_+(t)] + {\rm E}[dN_-(t)] -\sum_l {\rm E}[dN_l(t)]^2 \right\}  \del{\left\{\langle dN_+(t)\rangle + \langle dN_-(t)\rangle -\sum_l \langle dN_l(t)\rangle^2 \right\}}
\nonumber
\\ &\simeq \hbar^2\frac{\langle\hat{L}_+^\dag\hat{L}_+(t)\rangle+\langle\hat{L}_-^\dag\hat{L}_-(t)\rangle}{dt}
\nonumber
\\ &\simeq \delta(t'-t)\hbar^2\left[\Gamma\langle\hat{b}^\dag\hat{b}(t)\rangle+\Gamma'\langle\hat{b}\hat{b}^\dag(t)\rangle\right],
\end{align}
which reproduces $D^0(t)$ in Eqs.~(19)-(21) in the main text.
This indicates that the equal-time part of $D(t',t)$ is well understood by the Poisson process as discussed in the main text.

Next, we consider the nonlocal part $D^c(t',t)$.
For $t'> t$, \del{$D^c(t',t)$} the nonlocal part of ${\rm E}[\Delta I_s(t')\Delta I_s(t)]$ is calculated as follows.
If $dN_l=1$ is observed at time $t$, the quantum state of the system jumps to a post-measurement state. 
This quantum jump is described by a change in the density operator as
\begin{align}
\label{eq:app_rhos}
    \rho_s^l(t)=\frac{\hat{L}_l\rho_s(t)\hat{L}_l^\dag}{\langle\hat{L}_l^\dag\hat{L}_l\rangle}.
\end{align}
The term \del{$\langle dN_{l'}(t')dN_l(t)\rangle$} ${\rm E}[dN_{l'}(t')dN_l(t)]$ in Eq.~(\ref{eq:app_s}) means the probability of getting $dN_{l'}=1$ at time $t'$ in the condition of observation of $dN_{l}=1$ at time $t$, which is described by
\begin{align}
&{\rm E}[dN_{l'}(t')dN_l(t)] \del{\langle dN_{l'}(t')dN_l(t)\rangle}= {\rm Pr}\left[dN_{l'}(t')=1|dN_l(t)=1\right]\times {\rm E}[dN_l(t)] \del{\langle dN_l(t)\rangle},
\end{align}
where ${\rm E}[dN_l(t)]$ \del{$\langle dN_l(t)\rangle$} is the probability getting $dN_l=1$ at time $t$ and the conditional probability ${\rm Pr}\left[dN_{l'}(t')=1|dN_l(t)=1\right]$ is a probability getting $dN_{l'}=1$ at $t'$ in the condition of $dN_l=1$ at $t$.
Since the density operator after getting $dN_l=1$ at time $t$ is given by Eq.~(\ref{eq:app_rhos}), this conditional probability is expressed as
\begin{align}
& {\rm Pr}\left[dN_{l'}(t')=1|dN_l(t)=1\right]=\Tr\left\{\hat{L}_{l'}^\dag \hat{L}_{l'} e^{\mathcal{G}(t'-t)}\rho_s^{l}(t)\right\},
\end{align}
where $\mathcal{G}$ is the time evolution operator defined in Eq.~(4) in the main text.
Thus, the correlation function \del{$D(t',t)$} ${\rm E}[\Delta I_s(t')\Delta I_s(t)]$ given in Eq.~(\ref{eq:app_s}) is calculated for $t'>t$ as
\begin{widetext}
\begin{align}
\del{D(t',t)} \hbar^2{\rm E}[\Delta I_s(t')\Delta I_s(t)]&=\hbar^2\sum_{l,l'}l'l\Tr\left\{\hat{L}_{l'}^\dag\hat{L}_{l'}e^{\mathcal{L}(t'-t)}\left[\hat{L}_l\rho_s\hat{L}_l^\dag\right]\right\}
-\hbar^2\sum_{l,l'} l'l\langle \hat{L}_{l'}^\dag\hat{L}_{l'}(t')\rangle\langle \hat{L}_l^\dag\hat{L}_l(t)\rangle
\\\nonumber&= 
\hbar^2\int d^2\beta_1 P(\beta_1,\beta_1^*,t) \Tr\left\{(\Gamma\hat{b}^\dag\hat{b}-\Gamma'\hat{b}\hat{b}^\dag) e^{\mathcal{L}(t'-t)}(\Gamma\hat{b}\ket{\beta_1}\bra{\beta_1}\hat{b}^\dag-\Gamma'\hat{b}^\dag\ket{\beta_1}\bra{\beta_1}\hat{b})\right\}
\\\nonumber&\hspace{1cm}-\hbar^2(\Gamma\langle\hat{b}^\dag\hat{b}(t')\rangle-\Gamma'\langle\hat{b}\hat{b}^\dag(t')\rangle)(\Gamma\langle\hat{b}^\dag\hat{b}(t)\rangle-\Gamma'\langle\hat{b}\hat{b}^\dag(t)\rangle)
\\\nonumber&=
\hbar^2\int d^2\beta_2d^2\beta_1 P(\beta_2,\beta_2^*,t'|\beta_1,\beta_1^*,t) \left[\Gamma|\beta_2|^2-\Gamma'(|\beta_2|^2+1)\right] \\\nonumber&\hspace{2cm}\times\left[\Gamma|\beta_1|^2-\Gamma'(|\beta_1|^2-\partial_{\beta_1}\beta_1-\partial_{\beta_1^*}\beta_1^*+1+\partial_{\beta_1}\partial_{\beta_1^*})\right]P(\beta_1,\beta_1^*,t)
\\\nonumber&\hspace{1cm}-\hbar^2(\Gamma-\Gamma')^2|\beta_0|^4e^{-(\Gamma-\Gamma')(t'+t)}
\\\nonumber&=
\hbar^2\int d^2\beta_2d^2\beta_1d^2\beta \mathcal{N}(t'-t)^{-1} \exp\left(-\frac{\pi}{\mathcal{N}(t'-t)}\left|\beta_2-\beta_1 e^{-\left(\frac{\Gamma-\Gamma'}{2}+i\omega_0\right)(t'-t)}\right|^2\right)  \left[\Gamma|\beta_2|^2-\Gamma'(|\beta_2|^2+1)\right] 
\\\nonumber&\hspace{2cm}\times\left[\Gamma|\beta_1|^2-\Gamma'(|\beta_1|^2-\partial_{\beta_1}\beta_1-\partial_{\beta_1^*}\beta_1^*+1+\partial_{\beta_1}\partial_{\beta_1^*})\right]P(\beta_1,\beta_1^*,t|\beta,\beta^*,0^+)P(\beta,\beta^*,0^+)
\\\nonumber&\hspace{1cm}-\hbar^2(\Gamma-\Gamma')^2|\beta_0|^4e^{-(\Gamma-\Gamma')(t'+t)}
\\\nonumber&=
\hbar^2\int d^2\beta_1d^2\beta \left[(\Gamma-\Gamma')|\beta_1|^2-\Gamma'\right]e^{-(\Gamma-\Gamma')(t'-t)}
\\\nonumber&\hspace{2cm}\times\left[(\Gamma-\Gamma')|\beta_1|^2-\Gamma'(-\partial_{\beta_1}\beta_1-\partial_{\beta_1^*}\beta_1^*+1+\partial_{\beta_1}\partial_{\beta_1^*})\right]
\\\nonumber&\hspace{2cm}\times\mathcal{N}(t)^{-1}\exp\left(-\frac{\pi}{\mathcal{N}(t)}\left|\beta_1-\beta e^{-\left(\frac{\Gamma-\Gamma'}{2}+i\omega_0\right)t}\right|^2\right)\frac{\Gamma-\Gamma'}{\pi\Gamma'}\exp\left(-\frac{\Gamma-\Gamma'}{\Gamma'}\left|\beta-\beta_0\right|^2\right)
\\\nonumber&\hspace{1cm}-\hbar^2(\Gamma-\Gamma')^2|\beta_0|^4e^{-(\Gamma-\Gamma')(t'+t)}
\\\nonumber&=
\hbar^2 e^{-(\Gamma-\Gamma')(t'-t)} \int d^2\beta_1d^2\beta \left\{(\Gamma-\Gamma')^2|\beta_1|^4-4\Gamma'(\Gamma-\Gamma')|\beta_1|^2-\Gamma\Gamma'+2\Gamma'^2 \right\}
\\\nonumber&\hspace{2cm}\times\mathcal{N}(t)^{-1}\exp\left(-\frac{\pi}{\mathcal{N}(t)}\left|\beta_1-\beta e^{-\left(\frac{\Gamma-\Gamma'}{2}+i\omega_0\right)t}\right|^2\right)\frac{\Gamma-\Gamma'}{\pi\Gamma'}\exp\left(-\frac{\Gamma-\Gamma'}{\Gamma'}\left|\beta-\beta_0\right|^2\right)
\\\nonumber&\hspace{1cm}-\hbar^2(\Gamma-\Gamma')^2|\beta_0|^4e^{-(\Gamma-\Gamma')(t'+t)}
\\\nonumber&=
\hbar^2 e^{-(\Gamma-\Gamma')(t'-t)} \int d^2\beta_1' d^2\beta \left\{(\Gamma-\Gamma')^2(|\beta_1'|^4+4|\beta_1'|^2|\beta|^2e^{-(\Gamma-\Gamma')t}+|\beta|^4e^{-2(\Gamma-\Gamma')t})\right.
\\\nonumber&\hspace{4cm}\left.-4\Gamma'(\Gamma-\Gamma')(|\beta_1'|^2+|\beta|^2e^{-(\Gamma-\Gamma')t})-\Gamma\Gamma'+2\Gamma'^2 \right\}
\\\nonumber&\hspace{2cm}\times\mathcal{N}(t)^{-1}\exp\left(-\frac{\pi}{\mathcal{N}(t)}\left|\beta_1'\right|^2\right)\frac{\Gamma-\Gamma'}{\pi\Gamma'}\exp\left(-\frac{\Gamma-\Gamma'}{\Gamma'}\left|\beta-\beta_0\right|^2\right)
\\\nonumber&\hspace{1cm}-\hbar^2(\Gamma-\Gamma')^2|\beta_0|^4e^{-(\Gamma-\Gamma')(t'+t)}
\\\nonumber&=
\hbar^2 e^{-(\Gamma-\Gamma')(t'-t)} \int d^2\beta \left\{(\Gamma-\Gamma')^2\left(2\left(\frac{\mathcal{N}(t)}{\pi}\right)^2+4\frac{\mathcal{N}(t)}{\pi}|\beta|^2e^{-(\Gamma-\Gamma')t}+|\beta|^4e^{-2(\Gamma-\Gamma')t}\right)\right.
\\\nonumber&\hspace{2cm}\left.-4\Gamma'(\Gamma-\Gamma')\left(\frac{\mathcal{N}(t)}{\pi}+|\beta|^2e^{-(\Gamma-\Gamma')t}\right)-\Gamma\Gamma'+2\Gamma'^2 \right\}\frac{\Gamma-\Gamma'}{\pi\Gamma'}\exp\left(-\frac{\Gamma-\Gamma'}{\Gamma'}\left|\beta-\beta_0\right|^2\right)
\\\nonumber&\hspace{1cm}-\hbar^2(\Gamma-\Gamma')^2|\beta_0|^4e^{-(\Gamma-\Gamma')(t'+t)}
\\\nonumber&=
\hbar^2 e^{-(\Gamma-\Gamma')(t'-t)} \left\{(\Gamma-\Gamma')^2\left[2\left(\frac{\mathcal{N}(t)}{\pi}\right)^2+4\frac{\mathcal{N}(t)}{\pi}\left(|\beta_0|^2+\frac{\Gamma'}{\Gamma-\Gamma'}\right)e^{-(\Gamma-\Gamma')t}\right.\right.
\\\nonumber&\hspace{2cm}\left.\left.+\left(|\beta_0|^4+4|\beta_0|^2\frac{\Gamma'}{\Gamma-\Gamma'}+2\left(\frac{\Gamma'}{\Gamma-\Gamma'}\right)^2\right)e^{-2(\Gamma-\Gamma')t}\right]\right.
\\\nonumber&\hspace{2cm}\left.-4\Gamma'(\Gamma-\Gamma')\left[\frac{\mathcal{N}(t)}{\pi}+\left(|\beta_0|^2+\frac{\Gamma'}{\Gamma-\Gamma'}\right)e^{-(\Gamma-\Gamma')t}\right]-\Gamma\Gamma'+2\Gamma'^2 \right\}
\\\nonumber&\hspace{1cm}-\hbar^2(\Gamma-\Gamma')^2|\beta_0|^4e^{-(\Gamma-\Gamma')(t'+t)}
\\\nonumber&=-\hbar^2\Gamma\Gamma'e^{-(\Gamma-\Gamma')(t'-t)}
.
\end{align}
\end{widetext}
By a lengthy calculation, we can derive the same result for $D^c(t',t)$ in the main text.
In short, $D^c(t',t)$ is related to the dynamics of the spin state after the quantum jump accompanied by a relaxation process. \del{We note that this nonlocal part of $D^c(t',t)$ appears even in thermal equilibrium because of the nature of Bose statistics in the spin system.}

\section{Experimental condition underlying the calculation}
\label{exp:condition}

In the main text, we propose a feasible experimental protocol to observe the Fano factor using an ultrafast pump and probe technique applied to a thin film of ferromagnetic permalloy sapmle. In this appendix, we describe the specific experimental conditions underlying our calculation.
 
First, in this study, we focus on the time period during which distinct spin precession is observed, typically occurring several tens of picoseconds after laser pulse excitation. During this relatively slower time period, equilibration between the electron and phonon reservoirs in metals is likely achieved~\cite{DelFatti2000}. Furthermore, micromagnetic simulations indicate that energy transfer from high-energy excitations in the spin system to lower-order spin-wave excitations, is expected during this period~\cite{Djordjevic2007}. 
It takes a finite amount of time for the spin-wave front to propagate through the entire thickness of the ferromagnetic layer, as the initial demagnetization and remagnetization occur within the penetration depth of the ultrafast laser pulse, typically 20--30 nm~\cite{Djordjevic2007}. In our calculation, we assume a permalloy film thickness of 5 nm, which is smaller than this penetration depth. As a result, uniform equilibration between the electron, spin, and phonon systems is likely, with the temperatures of these three systems becoming nearly equal~\cite{Kirilyuk2010}.

Second, we assume low-temperature conditions, sufficiently below the Curie temperature of the ferromagnet, where variations in magnitude of the magnetization due to small temperature changes can be neglected. The assumption of a fixed magnetization magnitude in the phenomenological Landau-Lifshitz-Gilbert (LLG) equation, when using a macrospin approach, can sometimes fail at higher temperatures near the Curie temperature, particularly in micromagnetic domain configurations~\cite{Chubykalo-Fesenko2006}. In this higher temperature range, the Landau-Lifshitz-Bloch (LLB) equation~\cite{Garanin1997}, which accounts for the longitudinal relaxation of the magnetization, is more appropriate for describing the macrospin system. Notably, the LLB equation reduces to the LLG equation in the zero-temperature limit~\cite{Chubykalo-Fesenko2006}, meaning the use of the LLG equation remains valid at sufficiently low temperatures~\cite{Nowak2005}.
In this study, we consider temperatures below 5 K (as shown in Fig. 2(b) of the main text), which are much lower than the Curie temperature of typical ferromagnets (approximately 550 K for body-centered-cubic permalloy~\cite{Yu2008}). Additionally, the effective temperature of the spin system several tens of picoseconds after laser-heat pulse excitation is not substantially elevated, owing to the equilibration between the electron, spin, and phonon systems~\cite{Pankratova2022}. Therefore, the assumption of a fixed magnetization magnitude in the LLG equation remains valid under these conditions. 
It has been observed that the dynamics of ultrafast demagnetization and recovery processes show no discernible correlation with ambient temperature, even down to $T$ = 1.6 K in ferromagnets~\cite{Xie2023}.

Third, we consider an optical pulse with a typical width of 1 ps, which is significantly shorter than the spin relaxation time. Assuming a Fourier transform-limited pulse in the visible frequency range, the corresponding spectral bandwidth is less than 1 nm (or approximately 2 meV). Within this narrow energy bandwidth of 2 meV, the variation in the frequency-dependent signal is minimal in typical ferromagnets \cite{Krinchik1968, Tikuisis2017}. Moreover, an optimal frequency region can be selected where this variation is negligible. Therefore, the experimental data from the Faraday rotation or Kerr effect signals can be readily translated into magnetization information using the complex dielectric tensor at a central frequency of the probe light.

Lastly, while fast acquisition of magnetization is essential to gather sufficient data for calculating the statistical properties of $C(t',t)$, a recent advancement in time-resolved magneto-optical Kerr effect (TR-MOKE) measurements has made this process more efficient. Specifically, a novel scheme utilizing two optical pulses with slightly different repetition rates has been demonstrated \cite{Nishikawa2023}, enabling a single TR-MOKE measurement in just 2.6 ms. This setup allows for the collection of a sufficient amount of data within a reasonable time frame to statistically derive $C(t',t)$.

All the above requirements are experimentally accessible, and our theoretical considerations can be validated accordingly.

\bibliography{Reference}
\clearpage